\documentclass{iaus}

\newcommand{\br}{_{\rm br}}

\usepackage{graphicx}

\title[GRB spectral modeling]{GRB spectral parameter modeling} % within the fireball model}

\author[Gregory D. Fleishman \& Fedor A. Urtiev]   %% give here short author list %%
{Gregory D. Fleishman$^{1,2}$
 \and Fedor A. Urtiev$^3$}

\affiliation{$^1$New Jersey Institute of Technology, Newark, NJ 07102, USA
 \\[\affilskip]
$^2$%Ioffe Physical-Technical Institute of the Russian Academy of
Ioffe Institute, St. Petersburg 194021, Russia \\ email: {\tt
gfleishm@njit.edu}
\\[\affilskip]
$^3$State Polytechnical University, St.Petersburg, 195251, Russia
\\email: {\tt zigzagworld@rambler.ru}}

\pubyear{2010}
\volume{274}  %% insert here IAU Symposium No.
\pagerange{119--126}
 \jname{Advances in Plasma Astrophysics }
\editors{A. Bonanno, E. de Gouveia Dal Pino \&  A. Kosovichev, eds.}
\begin{document}

\maketitle

\begin{abstract}
Fireball model of the gamma-ray bursts (GRBs) predicts generation of
numerous internal shocks, which  efficiently accelerate charged
particles and generate relatively small-scale stochastic magnetic
and electric fields. The accelerated particles diffuse in space due
to interaction with the random waves and so emit so called Diffusive
Synchrotron Radiation (DSR) in contrast to standard synchrotron
radiation they would produce in a large-scale regular magnetic
fields. In this contribution we present key results of detailed
modeling of the GRB spectral parameters, which demonstrate that the
non-perturbative DSR emission mechanism in a strong random magnetic
field is consistent with observed distributions of the Band
parameters and also with cross-correlations between them. %; this
%analysis allowed to restrict GRB physical parameters from the
%requirement of consistency between the model and observed
%distributions.

\end{abstract}

\firstsection % if your document starts with a section,
\section{Introduction}

The fireball model  of the gamma-ray burst (GRB) suggests that a
central engine produces a number of interacting relativistic shock
waves, whose  interactions, in the collisionless case, result in
generation of fluctuating electromagnetic fields and acceleration of
charged particles up to high energies. It is well established by now
that the magnetic and electric fields produced in the shock
interactions have often a significant random
component at various spatial scales. %The presence of the random
%component has a crucial effect on generation of the non-thermal
%emission from corresponding objects. Indeed, unlike regular gyration
%in the presence of a regular magnetic field,
As a result, the shock-accelerated charged particles moving through
a plasma with random electromagnetic fields experience random Lorenz
forces and so follow random paths representing a kind of spatial
diffusion. Accordingly, the particles produce a ``diffusive
radiation'' whose spectra depend on the type of the field (magnetic
or electric) and on spectral energy distribution of the field over
the spatial scales (Toptygin \& Fleishman 1987, Fleishman 2006,
Fleishman \& Toptygin 2007a, Fleishman \& Toptygin 2007b, Reville \&
Kirk 2010).

Individual spectra of the prompt GRB emission are typically well
fitted by a phenomenological Band function (Band et al. 1993), which
consists of low-energy (spectral index $\alpha$) and high-energy
(spectral index $\beta$) power-law regions smoothly linked at a
break energy $E\br$. The DSR was shown (Fleishman 2006) to produce
spectra consistent with those observed typically from the GRBs (Band
et al. 1993). It had yet been unclear, however, if the DSR spectra
are naturally consistent with observed \emph{distributions} of the
GRB spectral parameters  (Preece et al. 2000; Kaneko et al. 2006)
and what ranges of physical GRB parameters are needed to reconcile
the theoretical spectra with the observed ones. In this conference
contribution we present results of the modeling of GRB prompt
emission generation by DSR in relativistically expanding GRB jets
presented in greater detail by \cite{Fl_Urt_2010}.

\section{Model}

\begin{figure}
\includegraphics[height=2.0in]{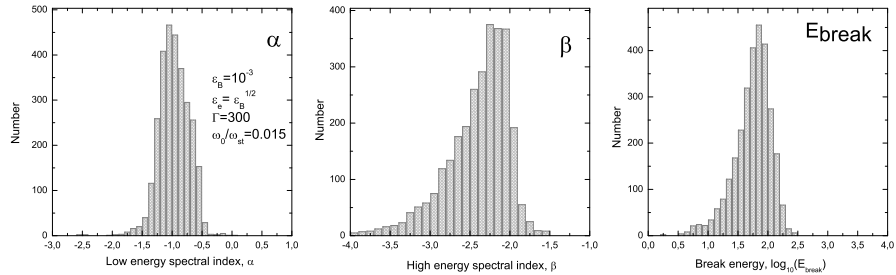}
\caption{\small Example of the model Band parameter ($\alpha$,
$\beta$, and $E\br$) distributions obtained within the DSR model
with strong random magnetic field.}\label{razd_3_3_hyst0903}
\end{figure}

\underline{\textbf{Formulation.}} Adopting a general internal
shocks/fireball concept we accept that a single binary collision of
relativistic internal shocks results in a single episode of the GRB
prompt emission (Fleishman \& Urtiev 2010). Microscopically, this
shock-shock interaction produces high levels of random magnetic
and/or electric fields and accelerates the charged particles up to
large ultrarelativistic energies; these particles interact with the
random fields to generate the gamma-rays. Although there are some
common general properties of all cases of relativistic shock
interactions, each shock-shock collision is, nevertheless, unique in
terms of combination of the physical parameters involved.
Accordingly, we adopt a set of standard ('mean') parameters
appropriate to account for the most global GRB properties, and then
consider if a reasonable scatter of those standard parameters is
capable of reproducing more detailed properties of the considered
class of events as a whole --- the statistical distributions of the
GRB spectral parameters and cross-correlations between them. To do
so, we considered a number of different emission models including
the standard synchrotron radiation and DSR regimes in case of either
weak or strong random magnetic field. The spectral slopes and breaks
depend on both the emission mechanism and combination of physical
parameters affecting the radiation spectra within a given mechanism.
Thus, the goal of the modeling is to establish if there exists a
parametric space making one or another theoretical model compatible
with the observational data on the GRB spectral properties.

\underline{\textbf{Results.}} \cite{Fl_Urt_2010} conclude that the
DSR model with the weak random magnetic field (jitter regime),
either perturbative or non-perturbative, cannot offer a consistent
fit to the observed $\alpha$ histogram. This complies with
independent criticisms of the jitter regime:
\cite{Kumar_McMahon_2008} noticed that it may imply an
unrealistically high level of inverse Compton emission, while
\cite{Kirk_Reville_2010} argued that the jitter case seems to be in
contradiction with the required high efficiency of the particle
acceleration at the shocks, so strong magnetic fluctuations are
needed to self-consistently accelerate electrons up to the gamma-ray
producing energies.

Thus, having the weak random field model (jitter regime) rejected,
we turn now to analysis of the strong random field case. According
to \cite{Fl_2005STR}, \cite{Fl_Biet_2007}, and
\cite{Reville_Kirk_2010_ar}, new asymptotes arise in this case,
which can yield broader $\alpha$ distribution. In this strong-field
regime, the model $\alpha$ distribution  depend on adopted $\nu$
distribution, which, within the adopted model, is straightforwardly
linked to the $\beta$ distribution, because $\beta=-\nu-1$. The
corresponding model results (Fig.~1) are in a remarkable agreement
with the observations. Indeed, the $\alpha$ histogram is a symmetric
one, it displays a peak at the right place, $\alpha=-1$, and its
bandwidth is comparable to that of the observed histogram. The
$\beta$ histogram almost repeats the observed one, displaying the
correct asymmetric shape and the peak at the right place,
$\beta=-2.2$. The $E\br$ histogram also agrees with the observed one
rather well displaying correct shape and bandwidth.
\begin{figure}
\centerline{\includegraphics[height=1.65in]{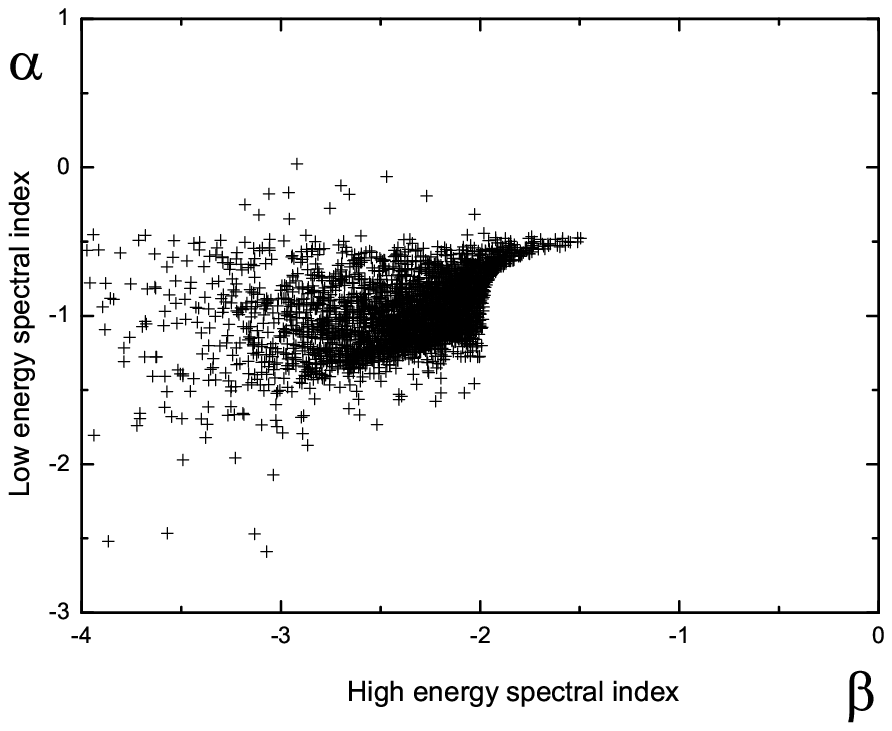}\hspace{-0.81cm}
\includegraphics[height=1.65in]{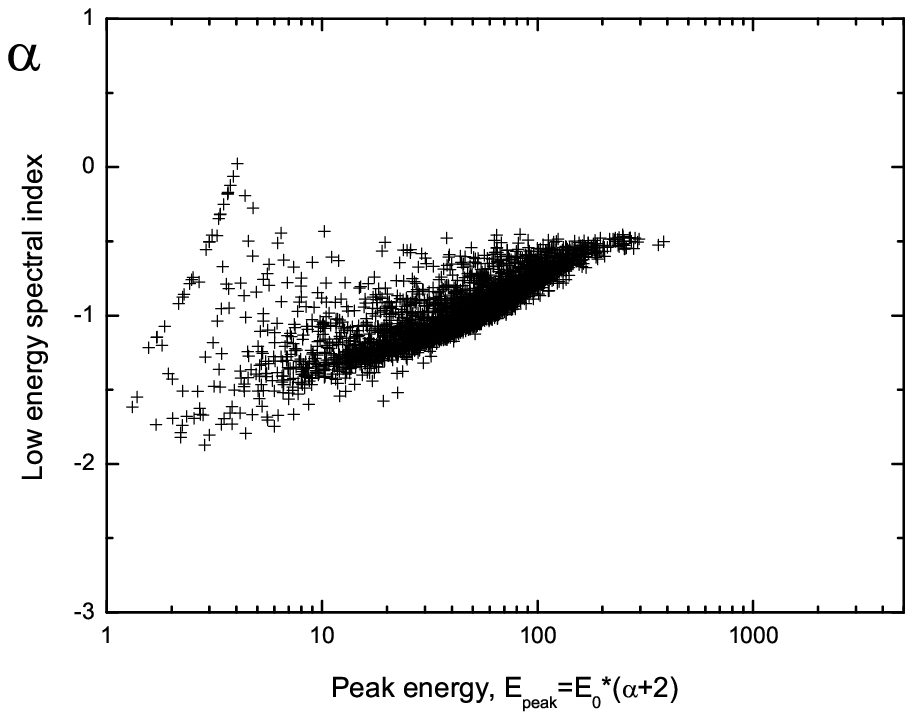}\hspace{-0.81cm}
\includegraphics[height=1.65in]{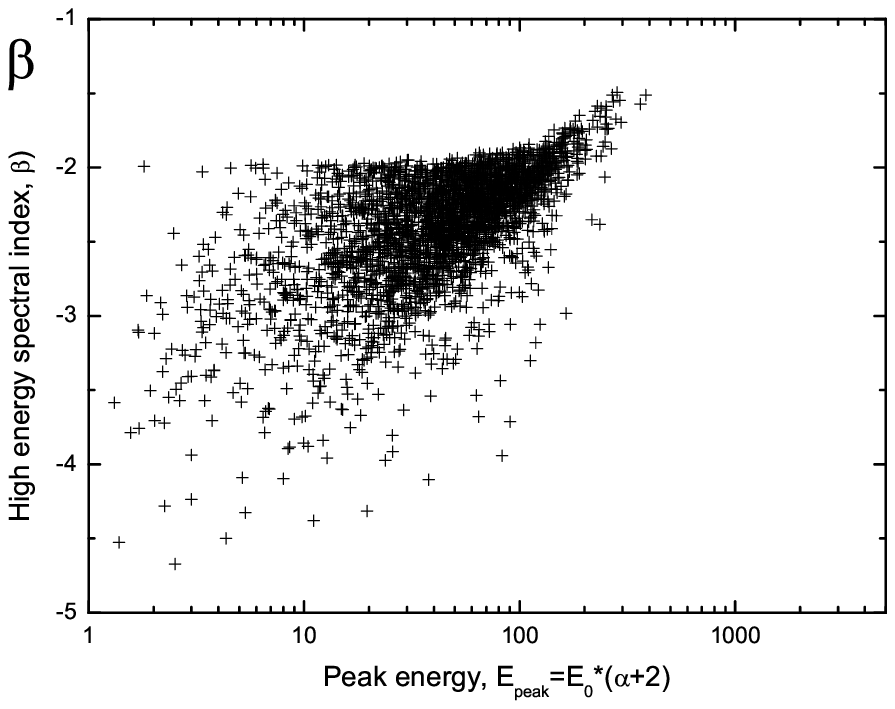}}
\caption{\small Cross-correlation of the model Band spectral
parameters $\alpha$ and $\beta$ (left);  $\alpha$ and $E_{\rm
peak}=(\alpha+2)E\br$ (middle); and $\beta$ and $E_{\rm
peak}=(\alpha+2)E\br$ (right).} \label{fig_correl}
\end{figure}

The cross-correlations between the spectral parameters derived from
the model (Fig.~2) are to be compared with fig. 31 from Kaneko et
al. (2006). Like in the observation, the spectral indices $\alpha$
and $\beta$ are not highly correlated, although in the model plot
the region of $-0.5<\alpha<0$ is underpopulated compared with the
observed plot (Kaneko et al. 2006). Two other plots are in
remarkable agreement with the observed cross-correlation plots,
presented in Kaneko et al. (2006).

\section{Conclusions}

We conclude that the developed model is naturally capable of
reproducing both the Band parameter histograms and their
cross-correlations, which is a remarkable success of the
non-perturbative DSR model in the presence of strong random magnetic
field.

%Our modeling shows that we can get an overall agreement between the
%model and observed histograms of the Band GRB spectral parameters
%within the non-perturbative DSR model with strong random magnetic
%field when ?0/?st ? 0.015. The histogram of the ?0/?pe ratio
%obtained for the model with ?0/?st = 0.015 is show in the last
%Figure. The distribution has a symmetric bell shape with the peak
%about 0.5, thus, Lmax ? 4?lsc and the required random field
%correlation length is indeed of the order of 10 plasma skin scales,
%which agrees with the idea of the random field generation by a
%two-stream instability in the internal shock interactions.

%\section*{Acknowledgments}

\textbf{Acknowledgments.} This work was supported in part by the
Russian Foundation for Basic Research, grants No. 08-02-92228,
09-02-00226, 09-02-00624. We have made use of NASA's Astrophysics
Data System Abstract Service.

%((\cite{Band1993}) \cite{Precee_2000,Kaneko_etal_2006})
%(\cite{Topt_Fl_1987,Fl_2006a,Fl_Topt_2007a,Fl_Topt_2007b}).

\bibliographystyle{apj} %\bibliography{DSR_PWNs,DSR_Langmuir,grb,fleishman}
\bibliography{DSR_PWNs,grb,fleishman}

\begin{thebibliography}{12}
\expandafter\ifx\csname natexlab\endcsname\relax\def\natexlab#1{#1}\fi

\bibitem[{{Band} {et~al. }(1993){Band}, {Matteson}, {Ford}, {Schaefer},
  {Palmer}, {Teegarden}, {Cline}, {Briggs}, {Paciesas}, {Pendleton}, {Fishman},
  {Kouveliotou}, {Meegan}, {Wilson}, \& {Lestrade}}]{Band1993}
{Band}, D. et al. 1993, \apj, 413, 281

\bibitem[{{Fleishman} (2005)}]{Fl_2005STR}
{Fleishman}, G.~D. 2005, ArXiv e-prints: astro-ph/0510317

\bibitem[{{Fleishman }(2006)}]{Fl_2006a}
---. 2006, \apj, 638, 348

\bibitem[{{Fleishman} \& {Bietenholz }(2007)}]{Fl_Biet_2007}
{Fleishman}, G.~D. \& {Bietenholz}, M.~F. 2007, \mnras, 376, 625

\bibitem[{{Fleishman} \& {Toptygin} (2007{\natexlab{a}})}]{Fl_Topt_2007a}
{Fleishman}, G.~D. \& {Toptygin}, I.~N. 2007{\natexlab{a}}, \mnras, 381, 1473

\bibitem[{{Fleishman} \& {Toptygin} (2007{\natexlab{b}})}]{Fl_Topt_2007b}
---. 2007{\natexlab{b}}, \pre, 76, 017401

\bibitem[{{Fleishman} \& {Urtiev} (2010)}]{Fl_Urt_2010}
{Fleishman}, G.~D. \& {Urtiev}, F.~A. 2010, \mnras, 406, 644

\bibitem[{{Kaneko} {et~al.} (2006){Kaneko}, {Preece}, {Briggs}, {Paciesas},
  {Meegan}, \& {Band}}]{Kaneko_etal_2006}
{Kaneko}, Y. et al. % , {Preece}, R.~D., {Briggs}, M.~S., {Paciesas}, W.~S., {Meegan},
%   C.~A., \& {Band}, D.~L.
  2006, \apjs, 166, 298

\bibitem[{{Kirk} \& {Reville} (2010)}]{Kirk_Reville_2010}
{Kirk}, J.~G. \& {Reville}, B. 2010, \apjl, 710, L16

\bibitem[{{Kumar} \& {McMahon} (2008)}]{Kumar_McMahon_2008}
{Kumar}, P. \& {McMahon}, E. 2008, \mnras, 384, 33

\bibitem[{{Preece} {et~al.}(2000){Preece}, {Briggs}, {Mallozzi}, {Pendleton},
  {Paciesas}, \& {Band}}]{Precee_2000}
{Preece}, R.~D. et al. % , {Briggs}, M.~S., {Mallozzi}, R.~S., {Pendleton}, G.~N.,
  %{Paciesas}, W.~S., \& {Band}, D.~L.
  2000, \apjs, 126, 19

\bibitem[{{Reville} \& {Kirk} (2010)}]{Reville_Kirk_2010_ar}
    {Reville}, B. and {Kirk}, J.~G. 2010, ArXiv e-prints: astro-ph.HE/1010.0872; this conf. proc.

\bibitem[{{Toptygin} \& {Fleishman} (1987)}]{Topt_Fl_1987}
{Toptygin}, I.~N. \& {Fleishman}, G.~D. 1987, \apss, 132, 213

\end{thebibliography}

\end{document}